\documentclass[a4paper]{article}

\usepackage{local-amsthm}
\usepackage{natbib}
\usepackage{preamble}
\usepackage{newpxtext}
\usepackage{newpxmath}

\begin{document}

\title{Reflections on existential types}
\author{Jonathan Sterling}

\maketitle
\begin{abstract}
  Existential types are reconstructed in terms of \emph{small reflective
  subuniverses} and dependent sums. The folklore decomposition detailed here
  gives rise to a particularly simple account of first-class modules as a mode
  of use of traditional second-class modules in connection with the modal
  operator induced by a reflective subuniverse, leading to a semantic
  justification for the rules of first-class modules in languages like OCaml
  and MoscowML.
  Additionally, we expose several constructions that give rise to
  semantic models of ML-style programming languages with both first-class
  modules and realistic computational effects, culminating in a model that
  accommodates higher-order first-class recursive modules \emph{and}
  higher-order store.
\end{abstract}

\NewDocumentCommand\JdgBox{m}{\colorbox{black!8}{\ensuremath{#1}}}
\NewDocumentCommand\IsModal{m}{#1\ \textit{modal}}

\section{Introduction}

Ever since the landmark paper \emph{Abstract Types Have Existential Type} by
\citet{mitchell-plotkin:1985}, the existential type construct has occupied a
central place in the study of programming languages, abstraction, and
modularity. The syntactic side of existential types has long been fairly
well-understood, as they are characterized by the now-standard rules depicted
in \cref{fig:ex-types} --- notable for their \emph{scoped} elimination rule, in
which the value of the first component of an existential package is not allowed
to ``escape'':
\begin{mathpar}
  \ebrule[$\ExistsSymbol$-elim]{
    \hypo{\IsTp{C}}
    \hypo{u:\Exists{x:A}Bx}

    \hypo{x:A,y:Bx\vdash vxy : C}

    \infer3{
      \UnpackAs{u}{\gl{x,y}}{vxy}
      : C
    }
  }
\end{mathpar}

In light of the Curry--Howard correspondence, these rules could be seen as a
proof term assigment for the natural deduction rules of existential
quantification in intuitionistic first-order logic. Such a perspective, though
meritable, does not seem to lead to an explanation of much definitude; indeed,
the competing Brouwer--Heyting--Kolmogorov correspondence would have suggested
instead the rules of \emph{dependent sums} (sometimes called ``strong sums'')
rather than the rules for existentials (sometimes called ``weak sums'')
that we have given above. If we are to find answers, therefore, they must come
from semantics.

\begin{figure}
  \begin{adjustwidth}{-1cm}{-1cm}
    \begin{mathpar}
      \ebrule[$\ExistsSymbol$-form]{
        \hypo{\IsKind{A}}
        \hypo{x:A\vdash \IsTp{Bx}}
        \infer2{
          \IsTp{\Exists{x:A} Bx}
        }
      }
      \and
      \ebrule[$\ExistsSymbol$-intro]{
        \hypo{u:A}
        \hypo{v : Bu}

        \infer2{
          \Pack\,\gl{u,v} : \Exists{x:A}Bx
        }
      }
      \and
      \ebrule[$\ExistsSymbol$-elim]{
        \hypo{\IsTp{C}}
        \hypo{u:\Exists{x:A}Bx}

        \hypo{x:A,y:Bx\vdash vxy : C}

        \infer3{
          \UnpackAs{u}{\gl{x,y}}{vxy}
          : C
        }
      }
      \and
      \ebrule[$\ExistsSymbol$-comp]{
        \hypo{\IsTp{C}}

        \hypo{u:K}
        \hypo{v : Au}

        \hypo{x:A,y:Bx\vdash wxy : C}

        \infer4{
          \UnpackAs{\prn{\Pack\,\gl{u,v}}}{\gl{x,y}}{wxy}
          \equiv wuv : C
        }
      }
      \and
      \ebrule[$\ExistsSymbol$-ext]{
        \hypo{\IsTp{C}}
        \hypo{z:\Exists{x:A}{Bx}\vdash uz,vz : C}
        \hypo{x:A,y:Bx\vdash u\,\prn{\Pack\,\gl{x,y}} \equiv v\,\prn{\Pack\,\gl{x,y}}:C}
        \infer3{
          z : \Exists{x:A}Bx\vdash uz \equiv vz : C
        }
      }
    \end{mathpar}
  \end{adjustwidth}

  \caption{The formation, introduction, elimination, computation, and
  (oft-overlooked) extensionality rules for existential types.}
  \label{fig:ex-types}

\end{figure}

\subsection{Existential types \vs dependent sums}\label{sec:ex-vs-sigma}

The difference between existential types and dependent sums is located in the
scoped elimination rule that we pointed out above.  Existentials are
counterposed to \emph{dependent sums}, which tend to be written using the
$\sum$-symbol; the latter have more flexible unscoped \emph{projection}-style
eliminators:
\begin{mathpar}
  \ebrule[$\sum$-elim(1)]{
    \hypo{u:\Sum{x:A}{Bx}}
    \infer1{u.1:A}
  }
  \and
  \ebrule[$\sum$-elim(2)]{
    \hypo{u:\Sum{x:A}{Bx}}
    \infer1{u.2 : B\,\prn{u.1}}
  }
\end{mathpar}

In comparison to dependent sums, the existential type interface suffers from
some obvious practical disadvantages when it comes to hierarchically structured
modular programming, as \citeauthor{macqueen:1986} argued forcefully in
\citeyear{macqueen:1986}:
\begin{quote}
  It appears that when building a collection of interrelated abstractions, the
  lower the level of the abstraction, the wider the scope in which it must be
  opened. We thus have the traditional disadvantages of block structured
  languages where low-level facilities must be given the widest visibility.
  --- \citet[p.~279]{macqueen:1986}
\end{quote}

On the other hand, the disadvantage of dependent sums is that it is not consistent
for $\Sum{\alpha:\KindTp}{A\alpha}$ to be a type, should $\alpha$ range
over \emph{all} types, by a variant of Girard's paradox~\citep{girard:1972}. It
is for this reason that \citet{macqueen:1986} and \citet{mitchell-plotkin:1985}
refer to Martin-L\"of's theory of \emph{type
universes}~\citep{martin-lof:itt:1975,martin-lof:1979} in their respective
discussions of dependent sums. Today we might employ the vocabulary of modern
module systems~\citep{stone-harper:2000} to summarize the situation as follows:
the dependent sum of a type family indexed in a kind is not in fact a type, but
a \textbf{module signature}. In particular, we have a weaker (but consistent)
formation rule for sums:
\begin{mathpar}
  \ebrule[$\sum$-form (restricted)]{
    \hypo{\IsKind{A}}
    \hypo{x:A\vdash\IsTp{Bx}}
    \infer2{
      \IsSig{\Sum{x:A} Bx}
    }
  }
\end{mathpar}

Such dependent sums that live in the ``higher'' universe of signatures are
adequate and, in fact, ideal from the perspective of hierarchical and modular
programming ``in the large''. On the other hand, because these sums are
signatures and not types, it means that we cannot pass around their
implementations dynamically, \eg by choosing an implementation of a component
based on the phase of the moon~\citep{harper:2016}. To support such dynamism is
exactly the motivation of the so-called \emph{first-class modules}, which
\citet{harper:2016} has advocated to view as existential types in the context
of an explicit phase-splitting; complementary to the perspective of \opcit, we
will develop here a \textbf{modal} construct that derives both existential
types and first-class modules from second-class modules simultaneously.

\section{Types, impredicativity, and first-class modules}\label{sec:easy}

\subsection{Introduction to reflective subuniverses}

Our analysis begins with the concept of a \emph{reflective
subuniverse}, and the similarity of its rules to those of existentials.
Although reflective subuniverses were first made explicit in category theory
and homotopy type theory~\citep{rijke-shulman-spitters:2020,hottbook}, they had
already appeared covertly within the classical programming languages literature
by \citeyear{abadi:1999}, as the \emph{protection monads} of
\citeauthor{abadi:1999}'s \emph{dependency core calculus}.

From a syntactic point of view, a reflective subuniverse
of a given universe $\UU$ is specified by an additional form of judgment
\colorbox{black!8}{$\IsModal{A}$} such that every type $A:\UU$ has a ``best
approximation'' $\Mod{A}$ as a modal type; unfurling what this means, we have
a type operator $\Mod$ equipped with the following structure:
\begin{mathpar}
  \ebrule[$\Mod$-form]{
    \hypo{A:\UU}
    \infer1{\IsModal{\Mod{A}}}
  }
  \and
  \ebrule[$\Mod$-modal]{
    \hypo{a : A}
    \infer1{
      \eta_Aa : \Mod{A}
    }
  }
  \and
  \ebrule[$\Mod$-elim]{
    \hypo{\IsModal{C}}
    \hypo{u : \Mod{A}}

    \hypo{x:A\vdash v{x} : C}

    \infer3{
      \Kwd{bind}~x=u~\Kwd{in}~v{x} : C
    }
  }
\end{mathpar}

We additionally have a computation rule and
extensionality rule for $\Kwd{bind}$:
\begin{mathpar}
  \ebrule[$\Mod$-comp]{
    \hypo{\IsModal{C}}
    \hypo{a : A}
    \hypo{x:A\vdash v{x}:C}
    \infer3{
      \Kwd{bind}~x=\eta_Aa~\Kwd{in}~v{x} \equiv v{a} : C
    }
  }
  \and
  \ebrule[$\Mod$-ext]{
    \hypo{\IsModal{C}}
    \hypo{x:\Mod{A}\vdash u{x},v{x} : C}

    \hypo{x:A\vdash u\,\prn{\eta_Ax}\equiv v\,\prn{\eta_Ax} : C}

    \infer3{
      z:\Mod{A}\vdash u{z}\equiv v{z}:C
    }
  }
\end{mathpar}

In a reflective subuniverse as described above, the modal operator $\Mod$ is
called the \emph{reflector} and a type $\Mod{A}$ is called the
\emph{reflection} of the type $A$.

\begin{remark}[Comparison to monads]
  The rules described here are similar to those of Moggi's monadic
  metalanguage, with some crucial differences. First of all, the $\Kwd{bind}$
  rule targets an arbitrary modal type rather than a type of the form
  $\Mod{B}$; second of all, the extensionality law that we have imposed is not
  valid for an arbitrary monad. It follows from the rules above that $\Mod$ is
  a strong monad on $\UU$, in fact an \emph{idempotent} one in the sense that
  $\mu : \Mod\Mod\to \Mod$ is a natural isomorphism.
\end{remark}

\NewDocumentCommand\dcc{}{\textsc{dcc}}

\begin{remark}[Dependency core calculus]
  The dependency core calculus (\dcc) of \citet{abadi:1999} contains exactly
  the judgmental structure and rules that we have described above,
  parameterized in a set of security levels $\mathcal{L}$:
  \begin{enumerate}

    \item Our \JdgBox{\IsModal{A}} judgment corresponds to \dcc's \JdgBox{\text{$A$ is
      protected at $\ell$}}.

    \item Our reflector $\Mod$ corresponds to \dcc's $T_\ell$.

  \end{enumerate}

  When first encountering the rules of \dcc, it is a rite of passage among
  programming language theorists to note their deviation from the rules of
  monads and try to ``fix'' them (see \citet{choudhury:2022} for a recent and
  creative example). Nonetheless, these rules are exactly the correct rules of
  a metalanguage for reflective subuniverses rather than arbitrary strong
  monads.
\end{remark}

We will say that a type $A:\UU$ is \emph{essentially modal} when the unit map
$\eta_A : A \to \Mod{A}$ is an isomorphism up to $\beta/\eta$-equivalence, \ie
it has a left and a right inverse. Of course, any modal type is essentially
modal, but the collection of essentially modal types has some remarkable
additional closure properties:
\begin{itemize}
  \item If $A$ and $B$ are essentially modal, then $A\times B$ is essentially
    modal.
  \item If $B$ is essentially modal, then $A\to B$ is essentially modal.
  \item If $B$ is essentially modal and $A$ is a retract of $B$, then $A$ is
    essentially modal.\footnote{A type $A$ is called a \emph{retract} of $B$ when there
    exists a section-retraction pair $\prn{s:A\to B,r : B \to A}$, \ie we have
    $r\circ s = \Con{id}_A$.}
\end{itemize}

The above justifies adding additional rules for judgmentally modal types, as
\citet{abadi:1999} have done for the sake of convenience:
\begin{mathpar}
  \ebrule{
    \hypo{\IsModal{A}}
    \hypo{\IsModal{B}}
    \infer2{
      \IsModal{A\times B}
    }
  }
  \and
  \ebrule{
    \hypo{A : \UU}
    \hypo{\IsModal{B}}
    \infer2{\IsModal{A\to B}}
  }
\end{mathpar}

Note that adding such rules does not change the expressive power of the
language, but it can make it more convenient to use.

\subsection{Existential types in a \emph{small} reflective subuniverse}\label{sec:intro:modal-ex}

We recall both the elimination rule for existential types and the elimination
rule for the reflection operator of a reflective subuniverse:
\begin{mathpar}
  \ebrule[$\ExistsSymbol$-elim]{
    \hypo{\IsTp{C}}
    \hypo{u:\Exists{x:A}Bx}

    \hypo{x:A,y:Bx\vdash vxy : C\strut}

    \infer3{
      \UnpackAs{u}{\gl{x,y}}{vxy}
      : C
    }
  }
  \and
  \ebrule[$\Mod$-elim]{
    \hypo{\IsModal{C}}
    \hypo{u : \Mod{A}}

    \hypo{x:A\vdash vx : C\strut}

    \infer3{
      \Kwd{bind}~x=u~\Kwd{in}~vx : C
    }
  }
\end{mathpar}

There is a superficial (syntactical) similarity in the structure of both rules.
Let us assume that we have a universe of signatures $\SIG$, as well as two
subuniverses $\TP,\KIND\subseteq\SIG$ such that $\TP:\KIND$; this is roughly
the configuration of ML-family programming languages.  Now \emph{additionally}
assume that $\TP$ is a reflective subuniverse of $\SIG$; as $\TP$ is already
assumed to be an element of $\SIG$, this amounts to saying that $\TP$ is a
\Alert{``small'' reflective subuniverse}. Unraveling these assumptions, we have
the following rules, as well as the computation and extensiality rules that we
specified earlier:
\begin{mathpar}
  \ebrule[$\Mod$-form]{
    \hypo{A:\SIG\strut }
    \infer1{\Mod{A}:\TP\strut}
  }
  \and
  \ebrule[$\TP$-modal]{
    \hypo{A:\TP\strut}
    \infer1{\IsModal{A}\strut}
  }
  \and
  \ebrule[$\Mod$-intro]{
    \hypo{\strut a : A}
    \infer1{
      \eta_Aa : \Mod{A}
      \strut
    }
  }
  \and
  \ebrule[$\Mod$-elim]{
    \hypo{\IsModal{C}\strut}
    \hypo{u : \Mod{A}}

    \hypo{x:A\vdash v{x} : C}

    \infer3{
      \strut
      \Kwd{bind}~x=u~\Kwd{in}~v{x} : C
    }
  }
\end{mathpar}

Now suppose in addition that $\SIG$ is closed under dependent sums of families
of types indexed in a kind:
\begin{mathpar}
  \ebrule[$\sum$-form (restricted)]{
    \hypo{A:\KIND}
    \hypo{x:A\vdash Bx:\TP\strut}
    \infer2{\Sum{x:A}{Bx}:\SIG\strut}
  }
  \and
  \ebrule[$\sum$-intro]{
    \hypo{u:A}
    \hypo{v:Bu\strut}
    \infer2{\prn{u,v} : \Sum{x:A}{Bx}\strut}
  }
  \and
  \ebrule[$\sum$-elim(1)]{
    \hypo{u:\Sum{x:A}{Bx}\strut}
    \infer1{u.1 : A\strut}
  }
  \and
  \ebrule[$\sum$-elim(2)]{
    \hypo{u:\Sum{x:A}{Bx}\strut}
    \infer1{u.2 : B\,\prn{u.1}\strut}
  }
  \and
  \ebrule[$\sum$-comp(1)]{
    \hypo{u:A}
    \hypo{v:Bu}
    \infer2{\prn{u,v}.1 \equiv u : A}
  }
  \and
  \ebrule[$\sum$-comp(2)]{
    \hypo{u:A}
    \hypo{v:Bu}
    \infer2{\prn{u,v}.2 \equiv v : Bu}
  }
  \and
  \ebrule[$\sum$-ext]{
    \hypo{u:\Sum{x:A}{Bx}}
    \infer1{u \equiv \prn{u.1,u.2}:\Sum{x:A}{Bx}}
  }
\end{mathpar}

Then we observe that the existential type $\Exists{x:A}{Bx}$ could be
represented as the ``synthetic connective'' $\Alert{\Mod{\Sum{x:A}{Bx}}}$
obtained by taking the reflection of the dependent sum. The introduction and
elimination forms of existentials are defined likewise by macro expansion:
\begin{align*}
  \Exists{x:A}{Bx} &\leadsto \Alert{\Mod{\Sum{x:A}{Bx}}}\\
  \Pack\,\gl{u,v} &\leadsto \Alert{\eta\Sub{\Sum{x:A}{Bx}}\prn{u,v}}\\
  \UnpackAs{u}{\gl{x,y}}{vxy} &\leadsto
  \Alert{\Kwd{bind}~z=u~\Kwd{in}~v\,\prn{z.1}\,\prn{z.2}}
\end{align*}

Evidence for our encoding is obtained by deriving the rules of existentials from
the rules of the reflection operator and the dependent sum:
\begin{mathpar}
  \ebrule[$\ExistsSymbol$-form]{
    \hypo{A:\KIND}
    \hypo{x:A\vdash Bx:\TP}
    \infer2{
      \Sum{x:A}{Bx}:\SIG
    }
    \infer1{
      \Mod{\Sum{x:A}{Bx}}:\TP
    }
    \infer[zigzag]1{
      \Exists{x:A}{Bx} : \TP
    }
  }
  \and
  \ebrule[$\ExistsSymbol$-intro]{
    \hypo{u:A}
    \hypo{v:Bu}
    \infer2{
      \prn{u,v} :
      \Sum{x:A}{Bx}
    }
    \infer1{
      \eta\Sub{\Sum{x:A}{Bx}}\prn{u,v} : \Mod{\Sum{x:A}{Bx}}
    }
    \infer[zigzag]1{
      \Pack\,\gl{u,v} : \Exists{x:A}{Bx}
    }
  }
  \and
  \ebrule[$\ExistsSymbol$-elim]{
    \hypo{C:\TP}
    \infer1{\IsModal{C}}

    \hypo{u:\Exists{x:A}{Bx}}
    \infer[zigzag]1{u:\Mod{\Sum{x:A}{Bx}}}

    \hypo{x:A,y:Bx\vdash vxy:C}
    \infer0{\ldots}
    \infer2{x:\Sum{x:A}{Bx}\vdash v\,\prn{x.1}\,\prn{x.2} : C}

    \infer3{
      \Kwd{bind}~x = u~\Kwd{in}~v\,\prn{x.1}\,\prn{x.2} : C
    }

    \infer[zigzag]1{
      \UnpackAs{u}{\gl{x,y}}{vxy} : C
    }
  }
\end{mathpar}

It is also possible to verify that our encoding satisfies both the computation
and extensionality rules of existential types.
\begin{align*}
  &\UnpackAs{\prn{\Pack\,\gl{u,v}}}{\gl{x,y}}{wxy}
  \\
  &\quad\leadsto
  \Kwd{bind}~x=\eta\Sub{\Sum{x:A}{Bx}}\prn{u,v}~\Kwd{in}~w\,\prn{x.1}\,\prn{x.2}
  \\
  &\quad\equiv
  w\,\prn{u,v}.1\,\prn{u,v}.2
  \\
  &\quad\equiv
  wuv
\end{align*}

For extensionality, we assume $C:\TP$ and $z:\Exists{x:A}{Bx}\vdash uz,vz : C$
such that
$x:A,y:Bx\vdash u\,\prn{\Pack\,\gl{x,y}} \equiv v\,\prn{\Pack\,\gl{x,y}} :C$ to
check that for each $w:\Exists{x:A}{Bx}$ we have $uw\equiv vw:C$. Under our
encoding, this follows immediately from the extensionality laws for $\Mod$ and
dependent sums.

\subsection{Universal types in a small reflective subuniverse}\label{sec:intro:modal-forall}

Just as existential types can be obtained by taking the reflection of the
(predicative) dependent sum of signatures, we observe here that universal types
arise in the same way from (predicative) dependent products of signatures. In
particular, we suppose that $\SIG$ is closed under dependent products of
families of types indexed in a kind, as below:
\begin{mathpar}
  \ebrule[$\prod$-form (restricted)]{
    \hypo{A:\KIND}
    \hypo{\strut x:A\vdash Bx:\TP}
    \infer2{
      \Prod{x:A}{Bx}:\SIG\strut
    }
  }
  \and
  \ebrule[$\prod$-intro]{
    \hypo{\strut x:A\vdash ux : Bx}
    \infer1{
      \lambda x. ux : \Prod{x:A}{Bx}\strut
    }
  }
  \and
  \ebrule[$\prod$-elim]{
    \hypo{\strut u : \Prod{x:A}Bx}
    \hypo{v:A}
    \infer2{ u\cdot v : Bv \strut}
  }
  \and
  \ebrule[$\prod$-comp]{
    \hypo{\strut x:A\vdash ux : Bx}
    \hypo{v:A}
    \infer2{
      \prn{\lambda x. ux}\cdot v \equiv uv : Bv
      \strut
    }
  }
  \and
  \ebrule[$\prod$-ext]{
    \hypo{\strut u:\Prod{x:A}{Bx}}
    \infer1{
      u \equiv \lambda x. u\cdot x : \Prod{x:A}{Bx}
      \strut
    }
  }
\end{mathpar}

Then we may close $\TP$ under universal types satisfying the following rules:
\begin{mathpar}
  \ebrule[$\ForallSymbol$-form]{
    \hypo{A:\KIND}
    \hypo{\strut x:A\vdash Bx:\TP}
    \infer2{
      \Forall{x:A}{Bx}:\TP\strut
    }
  }
  \and
  \ebrule[$\ForallSymbol$-intro]{
    \hypo{\strut x:A\vdash ux : Bx}
    \infer1{
      \Lambda x. ux : \Forall{x:A}{Bx}\strut
    }
  }
  \and
  \ebrule[$\ForallSymbol$-elim]{
    \hypo{\strut u : \Forall{x:A}Bx}
    \hypo{v:A}
    \infer2{ u\brk{v} : Bv \strut}
  }
  \and
  \ebrule[$\ForallSymbol$-comp]{
    \hypo{\strut x:A\vdash ux : Bx}
    \hypo{v:A}
    \infer2{
      \prn{\Lambda x. ux}\brk{v} \equiv uv : Bv
      \strut
    }
  }
  \and
  \ebrule[$\ForallSymbol$-ext]{
    \hypo{\strut u:\Forall{x:A}{Bx}}
    \infer1{
      u \equiv \Lambda x. u\brk{x} : \Forall{x:A}{Bx}
      \strut
    }
  }
\end{mathpar}

In particular, we define universal types via the following macro-expansions:
\begin{align*}
  \Forall{x:A}{Bx} &\leadsto \Alert{\Mod{\Prod{x:A}{Bx}}}\\
  \Lambda x. ux &\leadsto \Alert{\eta\Sub{\Prod{x:A}Bx}\prn{\lambda x. ux}}\\
  u\brk{v} &\leadsto \Alert{\Kwd{bind}~z=u~\Kwd{in}~z\cdot v}
\end{align*}

These macro-expansions do validate the rules of universal types:
\begin{mathpar}
  \ebrule[$\ForallSymbol$-form]{
    \hypo{A:\KIND}
    \hypo{x:A\vdash Bx:\TP}
    \infer2{\Prod{x:A}{Bx} : \SIG}
    \infer1{\Mod{\Prod{x:A}{Bx}} : \TP}
    \infer[zigzag]1{\Forall{x:A}{Bx}:\TP}
  }
  \and
  \ebrule[$\ForallSymbol$-intro]{
    \hypo{x:A\vdash ux : Bx}
    \infer1{
      \lambda x.ux : \Prod{x:A}{Bx}
    }
    \infer1{
      \eta\Sub{\Prod{x:A}{Bx}}\prn{\lambda x.ux} : \Mod{\Prod{x:A}{Bx}}
    }
    \infer[zigzag]1{
      \Lambda x. ux : \Forall{x:A}{Bx}
    }
  }
  \and
  \ebrule[$\ForallSymbol$-elim]{
    \hypo{x:A\vdash Bx:\TP}
    \hypo{v:A}
    \infer2{Bv:\TP}
    \infer1{\IsModal{Bv}}

    \hypo{u:\Forall{x:A}{Bx}}
    \infer[zigzag]1{u:\Mod{\Prod{x:A}{Bx}}}

    \infer0{z:\Prod{x:A}{Bx}\vdash z:\Prod{x:A}{Bx}}
    \hypo{v:A}
    \infer2{z:\Prod{x:A}{Bx}\vdash z\cdot v : Bv}
    \infer3{
      \Kwd{bind}~z=u~\Kwd{in}~z\cdot v : Bv
    }
    \infer[zigzag]1{
      u\brk{v} : Bv
    }
  }
\end{mathpar}

Next we check the computation rule:
\begin{align*}
  \prn{\Lambda x. ux}\brk{v}
  &\leadsto
  \Kwd{bind}~z=\eta\Sub{\Prod{x:A}{Bx}}\prn{\lambda x. ux}~\Kwd{in}~z\cdot v
  \\
  &\equiv
  \prn{\lambda x. ux}\cdot v
  \\
  &\equiv
  uv
\end{align*}

For the extensionality law, we fix $u:\Mod{\Prod{x:A}{Bx}}$ to check that
$u\equiv \Lambda x.u\brk{x}$ under our encoding. By the extensionality law for
$\Mod$, to check that $u \equiv \Lambda x.u\brk{x}$ for all $u$ is the same as
to check that for all $f:\Prod{x:A}{Bx}$ we have $\eta\Sub{\Prod{x:A}Bx}f
\equiv \Lambda x. \prn{\eta\Sub{\Prod{x:A}{Bx}}f} \brk{x}$. We proceed by
macro-expansion and computation:
\begin{align*}
  \Lambda x. \prn{\eta\Sub{\Prod{x:A}{Bx}}f}\brk{x}
  &\leadsto
  \eta\Sub{\Prod{x:A}{Bx}}\prn{
    \lambda x.
    \Kwd{bind}~z=\eta\Sub{\Prod{x:A}{Bx}}f~\Kwd{in}~z\cdot x
  }
  \\
  &\equiv
  \eta\Sub{\Prod{x:A}{Bx}}\prn{
    \lambda x.
    f\cdot x
  }
  \\
  &\equiv
  \eta\Sub{\Prod{x:A}{Bx}}f
\end{align*}

\subsection{Modal decomposition of first-class modules}\label{sec:modal-first-class}

In \cref{sec:intro:modal-ex} we have only explicitly assumed the closure of
$\SIG$ under dependent sums of the form $\Sum{x:A}Bx$ when $A:\KIND$ and
$x:A\vdash Bx:\TP$. This restriction echoes the explicit phase splitting of
\citet{harper-mitchell-moggi:1990}, but it is by no means forced.
\citet{sterling-harper:2021} have advocated to replace explicit phase
splittings (in which every signature is \emph{literally} the dependent sum of a
family of types indexed in a kind) with a \emph{synthetic} phase splitting, in
which the type theory of modules is simply Martin-L\"of type theory extended by
a (modal) phase distinction between compiletime and runtime data. In \opcit,
module signatures are naturally closed under arbitrary dependent sums and
products.

The modal reflection $\Mod$ can be thought of as a connective that
singlehandedly takes a module signature to the type of \emph{runtime packages}
of that signature. In the case of a signature of the form $\Sum{x:A}{Bx}$ where
$A$ is a kind and each $Bx$ is a type, the package type $\Mod\Sum{x:A}{Bx}$
encodes the existential type $\Exists{x:A}{Bx}$ as we have seen in
\cref{sec:intro:modal-ex}, but it is of course possible to apply the reflection
to any signature we want regardless of the way it was formed: the result is a
form of \emph{first-class modules}.

\begin{remark}
  The modal decomposition of the phase distinction by
  \citet{sterling-harper:2021} can be thought of as the first stage of a longer
  and more involved process of recasting \emph{every} last peculiarity of type
  theories for module systems as a completely orthodox modal extension of
  Martin-L\"of type theory.
  The present paper tackles a different aspect of modules, namely the
  relationship between types, modules, and impredicative polymorphism. Our
  approach can be viewed as a more modular alternative to the F-ing Modules
  tradition~\citep{rossberg-russo-dreyer:2014}, in which the peculiarities of
  module systems are reduced to the type structure of System~F via a highly
  non-trivial syntactic translation.
\end{remark}

\subsubsection{Existential and universal quantification over
\emph{modules}}\label{sec:quant-over-modules}

We now assume that signatures are closed under arbitrary dependent
sums and products, lifting the old restriction to families of types
indexed in a kind:
\begin{mathpar}
  \ebrule[$\sum$-form]{
    \hypo{A:\SIG}
    \hypo{x:A\vdash Bx:\SIG}
    \infer2{
      \Sum{x:A}{Bx}:\SIG
    }
  }
  \and
  \ebrule[$\prod$-form]{
    \hypo{A:\SIG}
    \hypo{x:A\vdash Bx:\SIG}
    \infer2{
      \Prod{x:A}{Bx}:\SIG
    }
  }
\end{mathpar}

With these more flexible rules in hand, we may use arguments identical to those
of \cref{sec:intro:modal-ex,sec:intro:modal-forall} to establish the closure of
$\TP$ under existential and universal types that take their indices in
signatures rather than only kinds:
\begin{mathpar}
  \ebrule[$\ExistsSymbol$-form (extended)]{
    \hypo{A:\SIG}
    \hypo{x:A\vdash Bx:\TP}
    \infer1{x:A\vdash Bx:\SIG}
    \infer2{\Sum{x:A}{Bx}:\SIG}
    \infer1{
      \Alert{\Mod{\Sum{x:A}{Bx}}}:\TP
    }
    \infer[zigzag]1{
      \Exists{x:A}{Bx} : \TP
    }
  }
  \and
  \ebrule[$\ForallSymbol$-form (extended)]{
    \hypo{A:\SIG}
    \hypo{x:A\vdash Bx:\TP}
    \infer1{x:A\vdash Bx:\SIG}
    \infer2{\Prod{x:A}{Bx}:\SIG}
    \infer1{
      \Alert{\Mod{\Prod{x:A}{Bx}}}:\TP
    }
    \infer[zigzag]1{
      \Forall{x:A}{Bx} : \TP
    }
  }
\end{mathpar}

A consequence of the closure under types of universal quantification over
elements of a given \emph{signature} is that we may write programs that take
modules as arguments without resorting (directly) to module functors; of
course, unraveling the encoding, such programs are ultimately packages of
module functors. We explore a concrete example of this phenomenon in
\cref{sec:programs-that-take-modules}

\subsubsection{Programs that take modules as arguments}\label{sec:programs-that-take-modules}

In Standard~ML and languages with similar ``second-class'' module systems, a
program that takes a module as an argument must be implemented as a
\emph{module functor}; thus such a program has a signature rather than a type,
and it cannot be (\eg) stored in the heap, \etc. To give an example, suppose we
have a module signature for an output interface, where \cd{$\alpha$\,eff}
is some ambient effect monad:
\begin{code}
  signature OUTPUT\_IO =
  sig
    type outstream
    val openOut : filepath -> outstream eff
    val write : oustream * string -> unit eff
    (* ... *)
  end
\end{code}

We can imagine a program that opens a file and writes ``Hello world'' to it,
and returns the open handle. Such a program in Standard ML must be written as a
module functor, taking a module \cd{O : OUTPUT\_IO} to a new module that
implements a function returning \cd{O.outstream eff}:
\begin{code}
  functor Hello (O : OUTPUT\_IO) :
  sig val hello : filepath -> O.outstream eff end =
  struct
    fun hello path =
      Eff.bind (O.openOut path) (fn h =>
      Eff.bind (O.write (h, "Hello world")) (fn \_ =>
      Eff.return h))
  end
\end{code}

Given our encoding of program-level universal quantification over
\emph{modules} from \cref{sec:quant-over-modules}, the \cd{hello} program could
be written much more succinctly without resorting to module functors:
\begin{code}
  fun hello (O : OUTPUT\_IO) (path : filepath) : O.outstream eff =
    Eff.bind (O.openOut path) (fn h =>
    Eff.bind (O.write (h, "Hello world")) (fn \_ =>
    Eff.return h))
\end{code}

\subsubsection{Programs that return modules}\label{sec:programs-that-return-modules}

In \cref{sec:programs-that-take-modules} we have used the fact that reflective
subuniverses are closed under dependent products to simplify the definition of
programs that take entire modules as arguments. On the other hand, it is also
possible to define programs that \emph{return} entire modules, perhaps even
computed using runtime inputs and/or computational effects.

\begin{figure}
  \begin{code}
    signature DATABASE\_IO =
    sig
      type db
      val openDb : filepath -> db eff
      val closeDb : db -> unit eff
      val execSql : db -> sql -> callback -> int eff
      (* ... *)
    end
    \-
    structure MockDatabaseIO : DATABASE\_IO = (*...*)
    structure RealDatabaseIO : DATABASE\_IO = (*...*)
  \end{code}
  \caption{A module signature for a database interface, together with two implementations.}
  \label{fig:database-io-sig}
\end{figure}

Consider for example the case of an interface to a database, whose signature we
depict in \cref{fig:database-io-sig}; can we write a program that will return
either a \emph{mocked} or \emph{real} database interface depending on the value
of an environment variable? It so happens that such a program can easily be
typed and implemented using the reflection operator $\Mod$, turning the module
signature \cd{DATABASE\_IO} into a first-class module type \Alert{\cd{$\Mod$ DATABASE\_IO}}:
\begin{code}
  val chooseDbInterface : (\Alert{$\Mod$ DATABASE\_IO}) eff =
    Eff.bind (getEnvFlag "USE\_MOCK\_DB")
    (fn true =>
        Eff.return ($\Alert{\eta\Sub{\cd{DATABASE\_IO}}}$ MockDatabaseIO)
      | false =>
        Eff.return ($\Alert{\eta\Sub{\cd{DATABASE\_IO}}}$ RealDatabaseIO))
\end{code}

Then a function opens and closes a given database, using either the mock or the
real database interface, can be written using a combination of the $\Kwd{bind}$
construct for $\Mod$ and the monadic bind of the ambient effect monad:
\begin{code}
  val main : unit eff =
    Eff.bind chooseDbInterface (fn dbPkg =>
    \Alert{bind DbInterface = dbPkg in}
    Eff.bind (DbInterface.openDb "mydb.sql") (fn db =>
    DbInterface.closeDb db))
\end{code}

\subsubsection{Modules with initialization effects}

A programming pattern identical to the one described in
\cref{sec:programs-that-return-modules} can be used to explain modules that
exhibit initialization effects, \eg binding an abstract type to a given runtime
state. For instance, we can consider the interface of a monotone symbol table
as follows:
\begin{code}
  signature SYMBOL\_TABLE =
  sig
    type symbol
    val fresh : symbol eff
    val symbolEq : symbol * symbol -> bool
  end
\end{code}

To allocate a new symbol table, we define a function that returns a
\emph{package} of type \Alert{\cd{$\Mod$ SYMBOL\_TABLE}} under the effect monad:
\begin{code}
  val initSymbolTable : (\Alert{$\Mod$ SYMBOL\_TABLE}) eff =
    Eff.bind (Ref.new 0) (fn r =>
    Eff.return
     (\Alert{$\eta\Sub{\cd{SYMBOL\_TABLE}}$}
      (struct
         type symbol = int
         val fresh =
           Eff.bind (Ref.get r) (fn i =>
           Eff.bind (Ref.set r (i + 1)) (fn \_ =>
           Eff.ret i))
         val symbolEq = Int.eq
       end)))
\end{code}

\subsubsection{Comparison to first-class modules in OCaml and MoscowML}

Both OCaml and MoscowML feature a form of first-class modules, and their rules
are similar in the broad strokes to those arising here through the encoding of
first-class modules as the \emph{reflections} of second-class modules; in
\cref{tab:rosetta-stone} we display a ``Rosetta stone'' that relates the
different notations for first-class modules to the modal account given here.
Our principled account of first-class modules in terms of a reflective
subuniverse can be seen, therefore, as a theoretical justification for the
rules of OCaml and MoscowML --- which were chosen on practical rather than
semantical grounds.

\begin{table}
  \begin{center}
    \begin{tabular}{lll}
      \toprule
      \textbf{This Paper} & \textbf{OCaml} & \textbf{MoscowML}\\
      \midrule
      $\Mod{A}$ & \cd{(module A)} & \cd{[A]}\\
      $\eta\Sub{A}u$ & \cd{(module u : A)} & \cd{[structure u as A]}\\
      $\Kwd{bind}~x=u$ & \cd{module x = (val u : A)} & \cd{structure x as A = u}
      \\
      \bottomrule
    \end{tabular}
  \end{center}
  \caption{A summary of the different notations for first-class modules compared with our modal account in terms of the reflective subuniverse.}
  \label{tab:rosetta-stone}
\end{table}

\subsubsection{Comparison to first-class modules in 1ML}

\citet{rossberg:2018} has argued that the so-called ``first-class'' modules of
OCaml and MoscowML should be referred to as \emph{packaged modules} rather than
first-class modules on the grounds that actual modules support a \emph{type
sharing} mechanism that is not expressible for the former without a
detour through core-level polymorphism. The example of \opcit involves a
signature containing an abstract type, and a function that aims to take two
implementations of that signature that share the same type component as in the
following code:
\begin{code}
  signature S = sig type t (* ... *) end
  fun example (X : S) (Y : S with type t = X.t) = (* ... *)
\end{code}

To encode the above in OCaml or MoscowML, it would be necessary take \cd{X} and
\cd{Y} as \emph{packaged} arguments, but then the dependency between the two
packages would not be expressible; \citet{rossberg:2018} explains that in
systems like OCaml and MoscowML, one must instead use an explicit universal
quantification over the type components of both modules and impose two type
sharing constraints in order to simulate the dependency:
\begin{code}
  fun example
    ($\alpha$ : type)
    (X : S with type t = $\alpha$)
    (Y : S with type t = $\alpha$) =
    (* ... *)
\end{code}

1ML's elaboration process ultimately addresses this defect of OCaml and
MoscowML's packaged modules, but we point out that our own account does not
suffer from this defect in the first place: we have observed in
\cref{sec:programs-that-take-modules} that the universe of types is
\emph{automatically} closed under dependent products of families over types
indexed in a signature as soon as the signature layer is closed under these
dependent products (this is a general fact about reflective subuniverses). Thus
the troublesome example depicted at the beginning of this section is already
expressible in our language without any arduous encoding.
Of course, it follows immediately from our analysis that the limitation of
OCaml and MoscowML's packaged modules criticized by \citet{rossberg:2018} is
only a superficial one, \ie a deficiency of syntax rather than of semantics.

\section{Semantic analysis of first-class modules}

So far we have investigated the purely syntactical aspects of existential
types, first-class modules, and their derivation from (small) reflective
subuniverses. Here we turn to semantics in order to justify to ourselves that
such scenarios are not merely a syntax-induced fantasy.

\subsection{A type theoretic metalanguage}

The metalanguage for our semantic investigations is intensional type theory
with dependent product types $\Prod{x:A}{Bx}$, dependent sum types
$\Sum{x:A}{Bx}$, and identification types $\Id{A}{u}{v}$ such that dependent
product types satisfy the function extensionality principle in relation to the
identification types. Although we use the vocabulary of univalent foundations
(\eg propositions, sets, \etc) we do not assume the univalence principle. Thus
our results are compatible with the \emph{usual} settings for models of
programming languages, as well as the future ones that we anticipate in the
world of univalent mathematics.

\begin{definition}[Propositions and sets]\label{def:propositions-and-sets}
  A type $A$ is called a \emph{proposition} when for every $x,y:A$ we may
  choose an element of $\Id{A}{x}{y}$.  A type $A$ is called a \emph{set} when
  for each $x,y:A$, the identification type $\Id{A}{x}{y}$ is a proposition.
\end{definition}

\begin{definition}[Contractibility]\label{def:contractibility}
  A type $A$ is \emph{contractible} when we have an element $a:A$ such that
  every other element $x:A$ can be identified with $a$. In other words, when we
  have an element of the type $\Sum{a:A}\Prod{x:A}\Id{A}{x}{a}$.
\end{definition}

\begin{definition}[Fibers]
  Given a function $f : A \to B$ and an element $b:B$, we will write
  $\Con{fib}_f b$ for the \emph{fiber} of $f$ at $b$ defined to be the dependent sum
  $\Con{fib}_f b\defeq \Sum{a:A}\Id{B}{fa}{b}$.
\end{definition}

\begin{definition}[Embeddings and equivalences]\label{def:emb+equiv}
  A function $f : A \to B$ is called an \emph{embedding} when each of its
  fibers is a proposition; it is called an \emph{equivalence} when each of its
  fibers is contractible.
\end{definition}

\begin{definition}[Universes]\label{def:universe}
  In this paper, a {universe} is simply a type $\UU$ equipped with a dependent
  type $A:\UU\vdash \Con{El}\Sub{\UU}A$. As a notational abuse, for each
  $A:\UU$ we will again abbreviate $\Con{El}\Sub{\UU}A$ by $A$.
\end{definition}

We do \emph{not} assume here that each universe is closed under the connectives
of type theory, as it will be useful for us to be able to make such assumptions
on a more granular level. Likewise, we do not assume that each universe is
univalent.

\begin{definition}[Closure under a type]\label{def:closure-under-types}
  A universe $\UU$ is said to be \emph{closed under} a type $A$ when
  there exists an element $\floors{A}:\UU$ and an equivalence $\alpha_A :
  \Con{El}\Sub{\UU}\floors{A}\simeq A$. As an abuse of notation, we will simply
  write $A\in\UU$ to assert that $\UU$ is closed under $A$, in this scenario we
  will write $A$ for $\floors{A}$ and leave the equivalence
  $\alpha_A:\Con{El}\Sub{\UU}\floors{A}\simeq A$ implicit.
\end{definition}

\NewDocumentCommand\Rst{mm}{\DelimMin{1}#1\Sub{\vert{#2}}}
\NewDocumentCommand\RstTiny{mm}{#1 \vert{#2}}

\begin{definition}[Reflection of a type]\label{def:reflection}
  A universe $\UU$ is said to \emph{reflect} a type $A$ when there exists an
  element $\Rst{A}{\UU}:\UU$ and a function $\eta_A : A \to \Rst{A}{\UU}$
  such that for every $B:\UU$ the map $B\Sup{\eta_{A}}$ determined by
  precomposition with $\eta_{A}$ depicted below is an equivalence:
  \begin{align}
    B\Sup{\eta_A} &: \prn{\Rst{A}{\UU}\to B}\to \prn{A\to B}\label{eqn:reflection-precmp}\\
    B\Sup{\eta_A}f &= f\circ\eta_A\notag
  \end{align}

  We will write $\Rec{\RstTiny{A}{\UU}}^B : \prn{A\to B}\to \prn{\Rst{A}{\UU}\to B}$ for the
  inverse to $B\Sup{\eta_A}$ so-determined; $\Rst{A}{\UU}$ is called the
  \emph{reflection} of $A$ in $\UU$, $\eta_A$ is called its
  \emph{unit}, and $\Rec{\RstTiny{A}{\UU}}$ is its \emph{recursion principle}.
\end{definition}

Note that the reflection of a type in a given universe is unique up to unique
isomorphism when it exists. It is not difficult to see that a universe reflects
any type $A$ that it is closed under, setting $\Rst{A}{\UU}\defeq A$ and
$\eta_A\defeq\Idn{A}$.

\begin{definition}[Subuniverses]\label{def:subuniverse}
  A \emph{subuniverse} $\UU\subseteq\VV$ is given by a predicate on
  $\VV$ that sends each $A:\VV$ to the proposition $\prn{A\in\UU}$,
  which is moreover \emph{replete} in the sense that when $f:A\to B$ is an equivalence
  where $A:\VV$ and $B\in \UU$, then $A\in\UU$.
\end{definition}

If $\VV$ were assumed to be univalent, then any predicate on $\VV$ would be replete
in the sense of \cref{def:subuniverse} and thus give rise to a subuniverse.

\begin{definition}[Reflective subuniverse]\label{def:reflective-subuniverse}
  A subuniverse $\UU\subseteq\VV$ is called \emph{reflective} in $\VV$
  when every $A:\VV$ is reflected in $\UU$.
\end{definition}

\begin{notation}
  In the case of a reflective subuniverse $\UU\subseteq\VV$, we will write
  $\Mod:\VV\to\UU$ for the operation that sends each type $A:\VV$ to its
  reflection $\Mod{A}\defeq \Rst{A}{\UU}$ in $\UU$; this operation is called
  the \emph{reflector}. In this scenario, we will often write
  $\VV\Sub{\Mod}\subseteq\VV$ for the reflective subuniverse for which $\Mod$
  is the reflector.
\end{notation}

A reflective subuniverse $\UU\Sub{\Mod}\subseteq\UU$ is closed under all
\emph{limit} type constructors that exist in $\UU$; for instance, if for some
$A,B:\UU\Sub{\Mod}$ we have $\prn{A\times B}\in \UU$, then it follows that
$\prn{A\times B}\in \UU\Sub{\Mod}$. This holds even for dependent products: for
any $A:\UU$ and $B : A\to \UU\Sub{\Mod}$ such that
$\prn{\Prod{x:A}Bx}\in\UU$, we furthermore have
$\prn{\Prod{x:A}Bx}\in\UU\Sub{\Mod}$. Likewise, given $A:\UU\Sub{\Mod}$ and
$a,b:A$, if it happens that $\prn{\Id{A}{a}{b}}\in\UU$, then so shall we also
have $\prn{\Id{A}{a}{b}}\in\UU\Sub{\Mod}$.

\subsection{Small reflective subuniverses and impredicativity}\label{sec:main}

Let $\UU,\VV$ be two universes such that $\UU\in\VV$ and $\VV$ is closed under
dependent products; note that we have not yet assumed that $\UU$ is reflective
in $\VV$ nor even that $\UU\subseteq\VV$.

\begin{definition}[Universal types]\label{def:universal-types}
  The nested universe $\UU\in\VV$ has \emph{universal types} when for all $A:\VV$
  and $B:A\to\UU$, we have $\prn{\Prod{x:A}{Bx}}\in \UU$.
\end{definition}

The following falls out directly from the results of
\citet{awodey-frey-speight:2018}.

\begin{restatable}[Universal types \vs{} reflectivity]{lemma}{LemUniversalVsReflective}\label{lem:universal-vs-reflective}
  Suppose that $\UU\in\VV$ and $\UU\subseteq\VV$ such every $A:\UU$ is a set
  and $\UU$ is moreover closed under identification types.
  Then $\UU\in\VV$ has universal types if and only if $\UU$ is reflective in $\VV$.
\end{restatable}

\begin{remark}
  Further observations of
  \citet{shulman:2017,shulman:blog:impredicative-encodings:3} concerning the
  splitting of idempotents in intensional type theory seem to suffice to
  generalize \cref{lem:universal-vs-reflective} to avoid the assumption that
  every $A:\UU$ is a set.
\end{remark}

\section{Concrete models of first-class modules}

In this section, we instantiate the results of the preceding sections to
produce several concrete semantic models of first-class modules.

\subsection{Models of total functional programming}\label{sec:in-realizability}

For any small subuniverse $\UU\in\VV$, \cref{lem:universal-vs-reflective}
establishes that closure under universal types is equivalent to $\UU$ being
reflective in $\VV$. Thus in light of our characterization of first-class
modules in terms of small reflective subuniverses
(\cref{sec:modal-first-class}), if we seek models of first-class modules, it
suffices to search for models of universal types.

Famously, \citet{reynolds:1984} observed that \textbf{\emph{``polymorphism is
not set-theoretic''}} by showing that the ``na\"ive'' model of universal types
in set theory, previously conjectured by \citet{reynolds:1983}, is ill-defined
due to insurmountable size issues. Category theorists would have anticipated
\citeauthor{reynolds:1984}' negative result, being aware of
\citeauthor{freyd:1964}'s 1964 observation in passing that any \emph{complete}
small category is necessarily a preorder~\citep{freyd:1964}.
\citeauthor{freyd:1964}'s observation pertains specifically to categories that
are defined using \emph{sets} as raw materials, \ie categories internal to the
(large) category of sets. Thus it did not contradict the results of
\citeauthor{freyd:1964} and \citeauthor{reynolds:1984} when
\citeauthor{pitts:1987} proclaimed that \textbf{\emph{``polymorphism is
set-theoretic, constructively''}}~\citep{pitts:1987}.

The meaning of \citeauthor{pitts:1987}' shocking slogan should be understood as
follows: when ``complete small category'' is interpreted in a \emph{different}
category of set-like objects, it is possible to find a complete small category
and, in particular, examples of non-trivial universes $\UU\in\VV$ closed under
universal types. The observations of \citet{pitts:1987} rely on the results of
\citet{hyland:1988}, who showed that categories of assemblies contain
non-trivial universes $\UU\in\VV$ closed under universal types.

\NewDocumentCommand\pca{}{\mathbb{A}}

\begin{assumption}
  We assume a \emph{partial combinatory algebra} $\pca$, which is a certain
  kind of computational model of untyped computation~\citep{van-oosten:2008}.
  Given $u,v\in \pca$ we will write $u\cdot v$ for the \emph{partial
  application} operator.
\end{assumption}

For example, $\pca$ could be the collection of Turing machines, or it could be
the syntax of $\lambda$-calculus under its $\alpha/\beta$-congruence, or it could be a
universal domain after Scott. It is not necessary to understand the details of
partial combinatory algebras in order to grasp the results of this section.

\begin{definition}
  A \emph{partial equivalence relation} is defined to be a relation
  $R\subseteq\pca\times\pca$ that is both symmetric and transitive.  A
  \emph{morphism} of partial equivalence relations $R\to S$ is given by an
  element $u\in\pca$ such that for all $x\mathrel{R}y$ we have both $u\cdot
  x,u \cdot y$ are defined and moreover $u\cdot x\mathrel{S} u\cdot y$.
\end{definition}

\NewDocumentCommand\GSec{}{\boldsymbol{\Gamma}}
\NewDocumentCommand\Rl{m}{\boldsymbol{E}\Sub{#1}}
\NewDocumentCommand\NEPow{m}{\mathcal{P}\Sub{\mathit{ne}}#1}

\begin{definition}
  An \emph{assembly} $X$ is defined to be a set $\GSec{X}$ together with a
  family of non-empty subsets $\Rl{X} : \GSec{X}\to \NEPow{\pca}$ sending each
  element of $\GSec{X}$ to its collection of ``realizers''.  A \emph{morphism}
  of assemblies $X\to Y$ is given by a function $f:\GSec{X}\to\GSec{Y}$ for
  which there exists an element $u\in\pca$ such that for each $x\in\GSec{X}$
  and $v\in \Rl{X}x$, $u\cdot v$ is defined and lies in $\Rl{Y}\prn{fx}$.
\end{definition}

\begin{definition}
  An assembly $X=\prn{\GSec{X},\Rl{X}}$ is called a \emph{modest set} when for
  each $x,y\in\GSec{X}$ if the intersection $\Rl{X}x\cap \Rl{X}y$ is inhabited,
  then $x=y$.
\end{definition}

Every partial equivalence relation $R$ gives rise to a modest set $M_R$: we
take $\GSec{M_R}$ to be the quotient of $\Compr{u\in\pca}{u\mathrel{R}u}$ by
the (total) equivalence relation induced by $R$. Then $\Rl{M_R}x$ is defined to
be the equivalence class $\Compr{v\in \pca}{\brk{v}_R=x}$. It turns out that
the mapping from partial equivalence relations to modest sets is an equivalence
of categories, whence we obtain a full embedding of PERs in assemblies.

\begin{construction}[Universes of assemblies~\citep{luo:1994}]\label{con:universe-of-assemblies}
  For each universe $V$ of the ambient set theory, we may define
  an assembly $\VV$ of $V$-small assemblies by taking $\GSec{\VV}$ to be the
  set of assemblies $X$ such that $\GSec{X}\in V$, and taking $\Rl{\VV}X$ to be
  all of $\pca$. The dependent assembly $X:\VV\vdash \Con{El}\Sub{\VV}X$ is
  defined to be $X$ itself.
\end{construction}

\begin{construction}[Universe of partial equivalence relations]
  We define an assembly $\UU$ of all partial equivalence relations by taking
  $\GSec{\UU}$ to be the set of partial equivalence relations and $\Rl{\UU}{R}$
  to be all of $\pca$. The dependent assembly $R:\UU\vdash \Con{El}\Sub{\UU}R$
  is defined to be the modest set $M_R$ determined by $R$.
\end{construction}

\begin{fact}
  It follows that $\UU\in\VV$ and $\UU\subseteq\VV$ for any $\VV$ built
  according to \cref{con:universe-of-assemblies}. Moreover, $\UU\in\VV$ is
  closed under universal types.
\end{fact}

\begin{corollary}\label{cor:realizability-modules}
  As $\UU$ is closed under equality types, it follows from
  \cref{lem:universal-vs-reflective} that $\UU\in\VV$ is a small reflective
  subuniverse and thus a model of first-class modules.
\end{corollary}

\paragraph*{Summary.\ }

Module signatures are modeled in $\bbrk{\SIG}\defeq\VV$ and types are modeled in
$\bbrk{\TP}\defeq\UU\in\VV$. First-class modules are accommodated using the
reflection $\Mod : \VV\to\UU$ which we have by
\cref{cor:realizability-modules}.

\subsection{Models with recursive types}\label{sec:sdt}

The results of \cref{sec:in-realizability} suffice to produce semantic models
of first-class modules in the setting of \emph{total functional programming};
in this section, we refine the model of \cref{sec:in-realizability} to account
for general recursion using the methods of \textbf{synthetic domain
theory}~\citep{hyland:1991}.

In the category of assemblies, we have a (univalent) universe $\mathbb{P}$
consisting of \emph{all} $\lnot\lnot$-closed propositions; we can isolate the
subuniverse $\Sigma\subseteq\mathbb{P}$ spanned by \emph{recursively
enumerable} propositions~\citep{rosolini:1986}, which we will use to define a
subuniverse of $\UU$ whose types support general recursion.  It so happens that
$\Sigma$ lies in $\UU$, so we may define a lifting monad $L:\UU\to\UU$ for
$\Sigma$-partial elements:
\[
  L A \defeq \Sum{p:\Sigma}\prn{p\to A}
\]

Let $\Con{F},\Con{I}$ be the \emph{final coalgebra} and \emph{initial
algebra} respectively for the endofunctor $L:\UU\to \UU$; these can be computed
as (co)inductive types. We should think of $\Con{I}$ as a type of
``generalized'' natural numbers, and $\Con{F}$ as the extension of
$\Con{I}$ by an infinite element. There is a canonical embedding $\iota :
\Con{I}\hookrightarrow\Con{F}$ that sends every ``finite generalized
natural number'' to itself, induced via Lambek's lemma by either the universal
property of the final coalgebra or the initial algebra.

\begin{definition}[\citet{longley:1995}]
  A type $A:\UU$ is called \emph{complete} if the precomposition map
  $A^\iota:A\Sup{\Con{F}}\to A\Sup{\Con{I}}$ is an equivalence. The type
  $A$ is called \emph{well-complete} when the lifted type $LA$ is complete.
\end{definition}

\NewDocumentCommand\UUWC{}{\UU\Sub{\textit{wc}}}

\begin{fact}
  The subuniverse $\UUWC\subseteq\UU$ of well-complete PERs is reflective in
  $\UU$, and moreover closed under $L$.
\end{fact}

\begin{corollary}
  As $\UU\in\VV$ is reflective and we have $\UUWC\in\VV$, the universe of
  well-complete PERs is a small reflective subuniverse of each $\VV$.
\end{corollary}

Well-completeness is a synthetic analogue of closure under suprema of
$\omega$-chains. In particular, the lifting $LA$ of any $A:\UUWC$ is closed
under general recursive definitions; furthemore, it even happens that $\UUWC$
is closed under recursive types.

\paragraph*{Summary.\ }

Module signatures are modeled in $\bbrk{\SIG}\defeq\VV$ and types are modeled in
$\bbrk{\TP}\defeq\UUWC\in\VV$.  First-class modules are accommodated via the
composite reflection $\VV\to\UU\to\UUWC$.
Moreover, we can model an effect monad on $\TP$
for general recursion, and we may even model recursive types of the form
$\mu\alpha. A\alpha$ for any $\alpha:\TP\vdash A\alpha:\TP$.

\subsection{Models with recursive \emph{modules}}\label{sec:sgdt}

An alternative to the methods of synthetic domain theory discussed in
\cref{sec:sdt} is given by \emph{synthetic \textbf{guarded} domain
theory}~\citep{bmss:2011} or SGDT, whose standard model takes place in the
\emph{topos of trees}, which is presheaves on the poset $\omega$ of natural
numbers with their standard order. SGDT centers around the use of type
connective $\Alert{\blacktriangleright}$ satisfying the rules of an
\emph{applicative functor} in the sense of \citet{mcbride-paterson:2008} to
stratify general recursive definitions in their finite unrollings.
\citet{palombi-sterling:2022} have shown that all the main results of SGDT
continue to apply when working in a different topos, such as a
\emph{realizability} topos into which the category of assemblies from
\cref{sec:in-realizability} embeds.

The model is constructed by replacing the poset of natural numbers from set
theory with the \emph{internal} poset $\omega$ of natural numbers from the
category of assemblies. Then we obtain a category of \emph{internal presheaves
of assemblies}, which is stratified by universes $\VV$ each obtained by taking
the Hofmann--Streicher lifting~\citep{hofmann-streicher:1997} of the
corresponding universe of ordinary assemblies. Then we have a subuniverse
$\UU\subseteq\VV$ of internal presheaves of PERs, which (as before) lies in
$\VV$ and is moreover reflective in $\VV$ as established by
\citet{sterling-gratzer-birkedal:2022}.
Every universe is closed under a \emph{guarded lifting monad}, which enables a
form of general recursive definition as a computational effect --- formally
quite different from the lifting monad $L$ from \cref{sec:sdt}, but achieving
much the same goal.

\begin{remark}[Recursive modules]
  In contrast to \cref{sec:sdt}, even the higher universes $\VV$ are closed
  under recursive types and recursive functions --- thus we model
  not only \emph{recursively defined modules} in the sense of
  \citet{dreyer:2007} but also recursively defined module signatures.
\end{remark}

\subsection{Models with recursive modules and higher-order store}

The guarded recursive model described in \cref{sec:sgdt} can be upgraded \`a la
\citet{sterling-gratzer-birkedal:2022} to a model with all the same formal
properties, that \emph{additionally} closes $\TP$ under an effect monad $T$
for higher-order store. The idea of \opcit is to iterate the presheaf
construction, defining a preorder $\mathbb{W}$ of semantic Kripke worlds
simultaneously with its collection of $\UU$-small co-presheaves by solving the
guarded recursive domain equation \Alert{$\mathbb{W} =
\prn{\mathbb{N}\rightharpoonup\Sub{\textit{fin}} {\blacktriangleright}\brk{\mathbb{W},\UU}}$}.

In particular, \opcit have shown how to model a type
constructor $\Con{ref} : \TP\to\TP$ for general (unrestricted) reference types.
Combined with the reflection, it becomes possible to store module packages in
the heap.

\subsection*{Acknowledgments}

I'm thankful to Robert Harper for many productive and enlightening
conversations on the subject of modules, existential types, and reflective
subuniverses; thanks to Dan Licata and Michael Shulman as well for helpful
discussions and suggestions. Thanks to El Pin Al for typographical corrections.
This work is funded by the European Union under the Marie Sk\l{}odowska-Curie
Actions Postdoctoral Fellowship project
\href{https://www.jonmsterling.com/typesynth.html}{\emph{TypeSynth: synthetic
methods in program verification}}. Views and opinions expressed are however
those of the authors only and do not necessarily reflect those of the European
Union or the European Commission. Neither the European Union nor the granting
authority can be held responsible for them.
 
\bibliographystyle{JFPlike}
\bibliography{references/refs-bibtex}

\clearpage
\appendix

\section{Appendix}

Let $\UU$ be a universe closed under dependent sums, function types, and
identification types such that every $A:\UU$ is a set in the sense of
\cref{def:propositions-and-sets}; let $\VV$ be another universe such that
$\UU\in\VV$ and moreover $\UU\in\VV$ is closed under universal types in the
sense of \cref{def:universal-types}.

\begin{lemma}\label{lem:warmup:main}
  Let $A$ be a type such that for each $C:\UU$, the function space $A\to C$
  lies in $\UU$; then $A$ is reflected by $\UU$.
\end{lemma}

The proof is an immediate application of the method of
\citet{awodey-frey-speight:2018}.

\begin{proof}
  First we define the ``wild'' reflection $\tilde{A}$ as follows:
  \[
    \tilde{A}\defeq\Prod{C:\UU}\prn{\prn{A\to C}\to C}
  \]

  The wild reflection lies in $\UU$ because $\UU\in\VV$ is assumed to have
  universal types, and moreover each $A\to C$ is assumed to lie in $\UU$.  We define a
  family of types indexed in $u:\tilde{A}$ governing ``naturality data'' for
  $u$:
  \begin{align*}
    \Con{isNatural} &: \tilde{A}\to \UU\\
    \Con{isNatural}\,u&\defeq
    \Prod{C,D:\UU}
    \Prod{f:C\to D}
    \Prod{h:A\to C}
    \Id{D}{u\,D\,\prn{f\circ h}}{f\,\prn{u\,C\,h}}
  \end{align*}

  Because each $D:\UU$ is assumed to be a set, it follows that each
  $\Con{isNatural}\,u$ is a proposition. Thus we define the actual reflection
  $\Rst{A}{\UU}$ to be the following subset:
  \[
    \Rst{A}{\UU}\defeq \Compr{u:\tilde{A}}{\Con{isNatural}\,u}
  \]

  The unit $\eta_A : A \to \Rst{A}{\UU}$ is defined to take $x:A$ to $\lambda
  C\,k. kx$; this function can be seen to be valued in $\Rst{A}{\UU}$ by
  definition. It remains to argue that for any $B:\UU$, the precomposition function
  $B\Sup{\eta_A} : \prn{\Rst{A}{\UU}\to B}\to \prn{A\to B}$ is an equivalence.
  Fixing $f : A \to B$, we must show that the fiber of $B\Sup{\eta_A}$ over $f$
  is contractible. First we inhabit this fiber by exhibiting the extension
  $f^\sharp\,u\defeq u\,B\,f$; to see that $f^\sharp$ does in fact extend $f$ along
  $\eta_A$, we compute:
  \[
    f^\sharp\,\prn{\eta_Aa} \equiv \eta_A a\,B\,f \equiv fa
  \]

  Thus the fiber of $B\Sup{\eta_A}$ over $f$ is inhabited by $\prn{f^\sharp,
  \Con{refl}}$. Next we fix any $\prn{h,p} :
  \Con{fib}\Sub{\prn{B\Sup{\eta_A}}}f$ to check that
  $\Id{\Con{fib}\Sub{\prn{B\Sup{\eta_A}}}f}{\prn{h,p}}{\prn{f^\sharp,\Con{refl}}}$.
  Because $B$ is assumed to be a set, it suffices to check only that
  $\Id{\Rst{A}{\UU}\to B}{h}{f^\sharp}$. Fixing $u:\Rst{A}{\UU}$, we must
  check: $h u = f^\sharp u$, or equivalently, $h u = u\,B\,f$. Because $u$ is
  assumed natural, we have $\Id{B}{u\,B\,f}{h\,\prn{u\,\Rst{A}{\UU}\,\eta_A}}$, so it
  suffices to check that $\Id{\Rst{A}{\UU}}{u\,\Rst{A}{\UU}\,\eta_A}{u}$, but this also follows
  from naturality.
\end{proof}

With this in hand, \cref{lem:universal-vs-reflective} is an immediate corollary.

\LemUniversalVsReflective*
 
\end{document}